\documentclass[referee]{raa}            

\usepackage{graphicx,times,amssymb}             
\input{epsf.sty}                        
\input{psfig.sty}                       
\begin{document}

   \title{The silicate model and carbon rich model of CoRoT-7b, Kepler-9d and Kepler-10b\,$^*$
\footnotetext{$*$ Supported by the National Natural Science Foundation of China.}
}

   \volnopage{Vol.0 (200x) No.0, 000--000}      
   \setcounter{page}{1}          

   \author{Yan-Xiang Gong
      \inst{1, 2}
   \and Ji-Lin Zhou
      \inst{1}
   }

   \institute{Department of Astronomy \& Key Laboratory of Modern Astronomy and Astrophysics in Ministry of Education, Nanjing University, Nanjing 210093, China; {\it zhoujl@nju.edu.cn}\\
        \and
             College of Physics and Electronic Engineering, Taishan University, Taian 271021, China\\
   }

   \date{Received 2011 month day; accepted 2011 month day}

\abstract{Possible bulk compositions of the super-Earth exoplanets, CoRoT-7b, Kepler-9d, and Kepler-10b are investigated by applying a commonly used silicate and a non-standard carbon model. Their internal structures are deduced using the suitable equation of state of the materials. The degeneracy problems of their compositions can be partly overcome, based on the fact that all three planets are extremely close to their host stars. By analyzing the numerical results, we conclude: 1) The iron core of CoRoT-7b is not more than $27\%$ of its total mass within 1 $\sigma$ mass-radius error bars, so an Earth-like composition is less likely, but its carbon rich model can be compatible with an Earth-like core/mantle mass fraction; 2) Kepler-10b is more likely with a Mercury-like composition, its old age implies that its high iron content may be a result of strong solar wind or giant impact; 3) the transiting-only super-Earth Kepler-9d is also discussed. Combining its possible composition with the formation theory, we can place some constraints on its mass and bulk composition.
}
\authorrunning{Y.-X. Gong \& J.-L. Zhou}            
\titlerunning{The silicate model and carbon rich model of CoRoT-7b, Kepler-9d and Kepler-10b}  

   \maketitle

%
%
\section{Introduction}           
\label{sect:intro}

Rocky planets (like the Earth in our solar system) located in the habitable zone of their stars are
presently the best candidates for harbouring extra-terrestrial life (L\'eger et al. 2011). Therefore, the search for rocky exoplanets plays an important role in the present exoplanet detections.
Recent observations of exoplanets have revealed 758 exoplanets (http://exoplanet.eu/, updated to 03 November 2011), mainly through stellar radial velocity measurements or photometric detection of planets transiting their host stars.
Among them,  more than 30 exoplanets are with minimum mass $<$ 10 Earth masses ($M_\oplus$).
They are called `super-Earths' and a faction of them may possibly be rocky planets.Recently,  the Kepler mission revealed 1235
planetary candidates with transit-like signatures detected during its first 4-month operation (Borucki et al.~\cite{borucki11} ).
Among them, 68 candidates of approximately Earth-sizes (${R_p} < 1.25\;{R_ \oplus }$), 288 super-Earth sizes
($1.25\;{R_ \oplus } < {R_p} < 2\;{R_ \oplus }$).  Beyond all doubt, we are entering the era of small exoplanets.
Some of them are expected to be terrestrial in nature.

 The understanding of super-Earths is of great interest to us due to their possible rocky nature.
 What is the bulk composition of these new planets? How were they formed?  These are all interesting yet puzzling questions.
  However, the information presently available based on observations is still limited for understanding their composition.
Through transit and radial velocity detections, planetary radii and masses are the only windows presently to investigate their
composition for most exoplanets. Among the detected super-Earths, their masses and the radii  show quite different signatures when compared to terrestrial planet of our solar system.
Their low average density tells us that they must primarily contain some light element which is lighter than water. They may have a significant gas envelope like Jupiter or Neptune in our solar system. However, super-Earths lacking of gas envelops, or `rocky super-Earths', may be also present in some cases, as we will discuss below.

 CoRoT-7b is the first super-Earth with a measured radius, $R = 1.68 \pm 0.09\;{R_ \oplus }$ (L\'eger et al.~\cite{leger09}). Its obtained mass combined with radial velocity measurements is $M = 4.8 \pm 0.8\;{M_ \oplus }$ (Queloz et al.~\cite{queloz09}). After 2009, the initial results have been revised by other authors. For example, after revision of CoRoT-7 stellar parameter by Bruntt et al. (\cite{bruntt10}), it yields $M = 5.2 \pm 0.8\;{M_ \oplus }$. Hatzes et al. (\cite{hatzes11}) performed another analysis of the data and got a larger mass, $M = 7.0 \pm 0.5\;{M_ \oplus }$. The works done by Boisse et al. (\cite{boisse11}) and Ferraz-Mello et al. (\cite{ferraz11}) got different results. In this paper, we do not consider those masses and use the initial data of CoRoT-7b only. Recently, two transiting super-Earths, Kepler-9d (Torres et al.~\cite{torres11}) and Kepler-10b (Batalha et al.~\cite{batalha11})  were discovered. Kepler-9d
was first discovered in 2010 as a planet candidate named KOI-377.03 (Holman et al.~\cite{holman10}). It was validated as a super-Earth and the reported radius is $R = 1.64_{-0.14}^{+0.19}\;{R_\oplus }$ (Torres et al.~\cite{torres11}). Although current spectroscopic observations are yet insufficient to establish its mass, a possible mass range can be found by Holman et al. (\cite{holman10}). The upper mass limit of $M = 7\;{M_ \oplus }$ corresponds to the maximum mantle-stripping limit (Marcus et al.~\cite{marcus10}) for a maximally iron-rich super-Earth. The lower mass limit is $\sim$3.5 $M_{\oplus}$ as a volatile-poor rocky planet with a Ganymede-like Fe/Si ratio. So we consider all possible compositions of Kepler-9d using this mass range in our paper. In this paper, we use Kepler-9d as an example to show what can be inferred for `transiting-only' hot super-Earths. Kepler-10b was discovered in early 2011, the reported mass and radius are ${M_p} = 4.56_{ - 1.29}^{ + 1.17}\;{M_ \oplus }$ and ${R_p} = 1.416_{ - 0.036}^{ + 0.033}\;{R_ \oplus }$ (Batalha et al.~\cite{batalha11}), respectively. The main parameters of these three exoplanets are summarized in Table 1.

\begin{table}
\begin{center}
\caption[]{Planet Parameters for Kepler-9d (Torres et al.~\cite{torres11}), Kepler-10b (Batalha et al.~\cite{batalha11}), and CoRoT-7b (L\'eger et al.~\cite{leger09}). The Ages of Their Host Stars are Also Listed Here.}\label{Tab:Parameter}


 \begin{tabular}{l c c c}
  \hline\noalign{\smallskip}
{\bf {Planet Name}} & {\bf {Kepler-10b}} & {\bf {Kepler-9d}} & {\bf {CoRoT-7b}}  \\
  \hline\noalign{\smallskip}
{Mass ($M_\oplus$)} & {$4.56_{ - 1.29}^{ + 1.17}$} & {$3.5 - 7.0$$^{(a)}$} & {$4.8 \pm 0.8$$^{(b)}$}   \\
 {Radius ($R_\oplus$)} & {$1.416_{ - 0.036}^{ + 0.033}$} & {$1.64_{-0.14}^{+0.19}$} & {$1.68 \pm 0.09$}  \\
 {Orbital Period (days)} & {$0.837495_{- 0.000005}^{+ 0.000004}$} & {$1.592851 \pm 0.000045$} & {$0.853585 \pm 0.000024 $}  \\
 {Equilibrium Temperature (K)} & {1833$^{(c)}$} & {$2026 \pm 60$} & {1800-2600}  \\
 {Orbital Semimajor Axis (AU)} & {$0.01684_{- 0.00034}^{+ 0.00032}$} & {$0.02730_{- 0.00043}^{+ 0.00042}$} & {$0.0172 \pm 0.00029$}  \\
 {Host Star Age (Ga)} & {$11.9 \pm 4.5$} & {$3.0 \pm 1.0$} & {$1.5_{- 0.3}^{+ 0.8}$$^{(d)}$}  \\
  \noalign{\smallskip}\hline
\end{tabular}
\begin{minipage}{13cm}
{\footnotesize  $^{(a)}$ Holman et al. (\cite{holman10}).\\
        $^{(b)}$ Queloz et al. (\cite{queloz09}).\\
        $^{(c)}$ Calculated value -- assuming a Bond albedo of 0.1 and a complete redistribution of heat for re-radiation.\\
        $^{(d)}$ http://www.exoplanet.eu/.}
\end{minipage}
\end{center}
\end{table}

The bulk composition of an exoplanet can not be uniquely determined by the measured mass and radius. When more than two chemical constituents are taken into consideration, a measured radius and mass correspond to more than one possible bulk composition, in other words, it is a degeneracy problem. The degeneracy problems of bulk compositions of low-mass exoplanets can be partly overcome if they are very close to their host stars because some chemical constituents may be ruled out a priori. Valencia et al. (2010) discussed the evolution of  close-in low-mass planets like CoRoT-7b. Due to the intense stellar irradiation and its small size, it is  unlikely to possess an envelope of hydrogen and helium of more than 1/10000 of its total mass. A relatively significant mass loss of  $\sim 10^{11}\;{\rm g~ s}^{- 1}$ is  expected and the result should prevail independently of the planet's composition, the hydrogen-helium gas envelope or water vapour atmosphere would escape within $\sim$1 Ga, which is shorter than the calculated age of 1.2-2.3 Ga for CoRoT-7b. The age of Kepler-9 is 3$\pm $1 Ga and Kepler-10 is a relatively old star (11.9$\pm$4.5 Ga), so volatile-rich solutions are less likely if the Kepler-9 and Kepler-10 planetary systems are old and evaporation has been substantial. Jackson et al. (\cite{jackson10}) studied the coupling of tidal evolution and evaporative mass loss on CoRoT-7b. Their investigation  manifests that the orbital decay caused by the star-planet tidal interaction enhanced its  mass loss rate. Such a large mass loss also suggests that, even if CoRoT-7b began as a gas giant planet, its original atmosphere has been completely evaporated now. As Kepler-9d and Kepler-10b are all small size close-in planets (with orbital periods 1.59 and 0.84 days, respectively), evaporation on the planets is serious due to high stellar irradiation. Thus, the origin of low-mass exoplanets, like Kepler-9d and CoRoT-7b cannot be inferred from the present observations: they may have always had a rocky composition; they may be remnants of a Uranus-like ice giant, or a gas giant with a small core that has been stripped of its gaseous envelope (Valencia et al.~\cite{valencia10}). The latest results from Leitzinger et al. (\cite{leitzinger11}) indicated that hydrogen-rich gas giants within the mass domain of Jupiter and Saturn cannot thermally lose such an amount of mass that CoRoT-7b and Kepler-10b would result in a rocky residue. They also concluded that these planets most likely were always rocky planets.

In this paper, we assume that CoRoT-7b, Kepler-9d, and Kepler-10b do not have significant gas envelopes and consider them as fully differentiated with all possible iron-to-silicate ratios. The differentiation assumption is not restrictive since all the terrestrial planets and some large satellites in the solar system are known to be differentiated. Using a relatively simple but well suited model for terrestrial planets in solar system, we discuss the interior structures of CoRoT-7b, Kepler-9d and Kepler-10b. We compare the silicate model and the carbon model (Seager et al. \cite{seager07}). Our main aims are: 1) infer their bulk composition and sharpen the observational constraints on their masses and radii, 2) compare their interior structures and infer some clues about their formation.

\section{Model and Method}
\label{sect:Model}
\subsection{Model Equations and Numerical Method}
If the rotation rate is not extreme, self-gravitating bodies are close to spherical. Spherical structures are certainly an adequate starting place for studies of exoplanets.
Three equations must be satisfied. They are the equation of hydrostatic equilibrium-Equation (\ref{eq:Hydro}), the mass conservation equation-Equation (\ref{eq:Mass}) and the equation of state (EOS)-Equation (\ref{eq:Eos}) for the material:
\begin{equation}\label{eq:Hydro}
\frac{{{\mathop{\rm d}\nolimits} P\left( r \right)}}{{{\mathop{\rm d}\nolimits} r}} =  - G\frac{{m\left( r \right)\rho \left( r \right)}}{{{r^2}}},
\label{eq:Hydro}
\end{equation}
\begin{equation}\label{eq:Mass}
\frac{{{\mathop{\rm d}\nolimits} m\left( r \right)}}{{{\mathop{\rm d}\nolimits} r}} = 4\pi {r^2}\rho \left( r \right),
\end{equation}
\begin{equation}\label{eq:Eos}
P\left( r \right) = f\left[ {\rho \left( r \right),  \;\;T\left( r \right)} \right].
\end{equation}
Where ${m\left( r \right)}$ is the mass contained within radius $r$, ${P\left( r \right)}$ is the pressure, ${T\left( r \right)}$ is the temperature, and ${\rho \left( r \right)}$ is the density of a spherical planet.
The thermal contributions to the conclusion can be ignored which has been testified in Seager et al. (\cite{seager07}) and this
approximation has been often used in other works (Fortney et al.~\cite{fortney07}; L\'eger et al.~\cite{leger04}). As a reasonable approximation, the temperature-independent EOS is also used in this work. Using variable transition, if pressure is a nonlinear function of density, Equation (\ref{eq:Hydro}) can be rewritten as,
\begin{equation}\label{eq:R1}
\frac{{{\mathop{\rm d}\nolimits} \rho }}{{{\mathop{\rm d}\nolimits} r}} =  - G\frac{{m\left( r \right)\rho \left( r \right)}}{{{r^2}}} \cdot \frac{1}{{{{{\mathop{\rm d}\nolimits} P\left( r \right)} \mathord{\left/
 {\vphantom {{{\mathop{\rm d}\nolimits} P\left( r \right)} {{\mathop{\rm d}\nolimits} \rho \left( r \right)}}} \right.
 \kern-\nulldelimiterspace} {{\mathop{\rm d}\nolimits} \rho \left( r \right)}}}}.
\end{equation}
We numerically integrate Equation (2) and Equation (4) starting at the planet's center $r = 0$ and using the inner boundary condition $m\left( 0 \right) = 0$
 and $\rho \left( 0 \right) = {\rho _c}$, where $\rho _c$ is a chosen central density. For the outer boundary condition, we use $P\left( {{R_p}} \right) = 0$. The choice of $\rho _c$ at the inner boundary and the outer boundary condition $P\left( {{R_p}} \right) = 0$ define the planetary radius $R_p$ and total mass ${M_p} = m\left( {{R_p}} \right)$. Integrating Equation (\ref{eq:Mass}) and Equation (\ref{eq:R1}) over and over for a range of $\rho _c$ provides the mass-radius relationship (see 3.1.1) for a given material. For differentiated model containing more than one kind of
material, we specify the desired fractional mass of the core and of each shell. We then integrate Equation (\ref{eq:Mass}) and Equation (\ref{eq:R1}) as specified
above, given a $\rho _c$ and outer boundary condition. Using the planetary mass, we switch from one material to the next where the desired fractional mass is
reached. Since we do not know the total mass that a given integration will yield ahead of
time, we generally need to iterate a few times (change the value of $\rho _c$) in order to produce a model with the desired distribution of material.

\subsection{The Choice of the EOS}
The materials used in this paper are ice VII, SiC, MgSiO$_{3}$ (enstatite, ab. en), (Mg$_{0.88}$, Fe$_{0.12}$)SiO$_{3}$ (perovskite, ab. pv), Fe($\alpha$), Fe($\varepsilon$). Ice VII is used to model the ice composition in large moons of giant planets. MgSiO$_{3}$ and (Mg$_{0.88}$, Fe$_{0.12}$)SiO$_{3}$ are used to model the silicate mantles of terrestrial planets or large moons in our solar system. Enstatite is a low-pressure mineralogical phase of silicate, so it is only used for the validation of the model (see section 3.1.2). Fe($\alpha$) and Fe($\varepsilon$) are different phases of iron under different pressure, Fe($\varepsilon$) is the phase occurring at higher pressures. We compare their differences in mass-radius relation for homogeneous planets. For CoRoT-7b, Kepler-9d, and Kepler-10b, we use Fe($\varepsilon$) to model their iron cores. Let us consider the pressure scope in our topic. $P_c$, the central pressure of the planet can be estimated as following
\begin{equation}
{P_c} \sim {{GM_p^2} \mathord{\left/ {\vphantom {{GM_p^2} {4R_p^4}}} \right.\kern-\nulldelimiterspace} {4R_p^4}}.
\end{equation}
It is about $~10^3$ GPa for the three exoplanets.  For $P \le 200$ GPa we use fits to experimental data, either the Vinet EOS (VE) Equation (\ref{eq:Vinet}) or the Birch-Murnaghan EOS (BME) Equation (\ref{eq:BME}). In general, when $P \ge {10^4}$ GPa, where electron degeneracy pressure becomes increasingly important, the Thomas-Fermi-Dirac (TFD) theoretical EOS (Salpeter \& Zapolsky~\cite{salpeter67}) must be employed, this is out of the range we need. The pressure range from approximately 200 to $10^3$ GPa is not easily accessible to experiment nor is it well described by the TFD EOS (Seager et al. \cite{seager07}). For all materials in this pressure regime we simply use the VE or BME.

For a derivation of these EOSs, see Poirier (\cite{poirier00}). The VE is
\begin{equation}\label{eq:Vinet}
P = 3{K_0}{\left( {\frac{\rho }{{{\rho _0}}}} \right)^{{2 \mathord{\left/
 {\vphantom {2 3}} \right.
 \kern-\nulldelimiterspace} 3}}}\left[ {1 - {{\left( {\frac{\rho }{{{\rho _0}}}} \right)}^{{{ - 1} \mathord{\left/
 {\vphantom {{ - 1} 3}} \right.
 \kern-\nulldelimiterspace} 3}}}} \right]\exp \left\{ {\frac{3}{2}\left( {{{K'}_0} - 1} \right)\left[ {1 - {{\left( {\frac{\rho }{{{\rho _0}}}} \right)}^{{{ - 1} \mathord{\left/
 {\vphantom {{ - 1} 3}} \right.
 \kern-\nulldelimiterspace} 3}}}} \right]} \right\},
\end{equation}
and the third-order finite strain BME is
\begin{equation} \label{eq:BME}
P = \frac{3}{2}{K_0}\left[ {{{\left( {\frac{\rho }{{{\rho _0}}}} \right)}^{{7 \mathord{\left/
 {\vphantom {7 3}} \right.
 \kern-\nulldelimiterspace} 3}}} - {{\left( {\frac{\rho }{{{\rho _0}}}} \right)}^{{5 \mathord{\left/
 {\vphantom {5 3}} \right.
 \kern-\nulldelimiterspace} 3}}}} \right]\left\{ {1 + \frac{3}{4}\left( {{{K'}_0} - 4} \right)\left[ {{{\left( {\frac{\rho }{{{\rho _0}}}} \right)}^{{2 \mathord{\left/
 {\vphantom {2 3}} \right.
 \kern-\nulldelimiterspace} 3}}} - 1} \right]} \right\}.
\end{equation}
Where ${\rho _0}$ is the reference density, ${K_0} =  - V{\left( {{{\partial P} \mathord{\left/ {\vphantom {{\partial P} {\partial V}}} \right. \kern-\nulldelimiterspace} {\partial V}}} \right)_{\,P = 0}}$ is the bulk modulus of the material, ${K'_0} = {\left( {{{\partial K} \mathord{\left/ {\vphantom {{\partial K} {\partial P}}} \right. \kern-\nulldelimiterspace} {\partial P}}} \right)_{\,P = 0}}$ is its pressure derivative. Their values for the materials used in this paper are summarized in Table 2. The VE is used for the $\varepsilon$-phase of iron because it is more suitable than the BME for extrapolation to high pressures (Poirier \cite{poirier00}). The third-order BME is used for other materials.

\begin{table}
\begin{center}
\caption[]{Parameters for the Birch-Murnaghan EOS (BME) or Vinet EOS (VE) Fits.}\label{Tab:Parameter}


 \begin{tabular}{l c c c c c}
  \hline\noalign{\smallskip}
{\bf {Materials}} & {${K_0}$ (GPa)} & {${K'_0}$} & {${\rho _0}$ (kg/m$^{3}$)}  & {\bf {EOS}}  & {\bf {References}} \\
\hline\noalign{\smallskip}
{ice VII} & {$23.7 \pm 0.9$} & {$4.15 \pm 0.07$} & {$1460$}  &  {BME}  & {1}  \\
{SiC} & {$227 \pm 3$} & {$4.1 \pm 0.1$} & {$3220$}  &  {BME}  & {2, 3}  \\
{MgSiO$_{3}$ (en)} &  {$125$} & {$5$} & {$3220$}  &  {BME}  & {2, 4}  \\
{(Mg$_{0.88}$, Fe$_{0.12}$)SiO$_{3}$ (pv)} & {$266 \pm 6$} & {$3.9 \pm 0.04$} & {$4260$}  &  {BME}  & {2, 5}  \\
{Fe($\alpha$)} & {$162.5 \pm 5$} & {$5.5 \pm 0.8$} & {$7860$}  &  {BME}  & {2, 6}   \\
{Fe($\varepsilon$)} & {$156.2 \pm 1.8$} & {$6.08 \pm 0.12$} & {$8300$}  &  {VE}  & {7}  \\
\noalign{\smallskip}\hline
\end{tabular}
\begin{minipage}{13cm}
{\footnotesize References: (1) Hemley et al. \cite{hemley87}, (2) Ahrens \cite{ahrens00}, (3) Alexsandrov et al. \cite{alexsandrov89}, (4) Olinger \cite{olinger77}, (5) Knittle \& Jeanloz \cite{knittle87}, (6) Takahashi \& Spain \cite{takahashi89}, (7) Anderson et al. \cite{anderson01}.}
\end{minipage}
\end{center}
\end{table}

\subsection{Carbon Model}
There is an assumption used in modeling the interior structure of solid exoplanets: their constituents are similar to the terrestrial planets in our solar
system. They have an iron core and a silicate envelope for the differentiated model. But based on current observation, other possibilities can not be ruled out. The facts tell us that there are large differences between extrasolar planets and planets in our solar system. For example, research about hot Jupiter tells us it is danger to assume some planets can only form in particular area. It is useful to take as broad a view as possible (Fortney et al. \cite{fortney07}). To other possible composition, Seager et al. (\cite{seager07}) presented the idea of carbon planets (see also Gaidos~\cite{gaidos00}). The planets in our solar system were born in an environment where the carbon-to-oxygen ratio C/O = 0.5. Their bulk compositions were determined by the high-temperature chemical equilibrium in protoplanetary disk. Silicates (i.e., Si-O compounds) are the dominant constituents in terrestrial mantles. But in an environment where C/O $\gtrsim$ 1, the condensation sequence changes dramatically (Larimer \cite{larimer75}; Lewis~\cite{lewis74}; Wood et al.~\cite{wood93}). As a result, the high-temperature condensates will be carbon-rich compounds. Kuchner \& Seager (\cite{kuchner06}) have explored some planet-formation scenarios and they proposed that carbon planets composed largely of SiC or other carbides should form in such environment. Carbon-rich nebula created by the disruptions of carbon-rich stars or white dwarfs is a natural cradle for carbon planets, so planets around pulsars and white dwarfs are candidates of carbon planets. For example, the planets around pulsar PSR1257+12 (Wolszczan et al.~\cite{wolszczan92}) may be carbon-rich exoplanets. In the well-known $\beta $ Pictoris debris disk (Roberge et al.~\cite{roberge06}), an exoplanet named $\beta$ Pic b was discovered by Lagrange et al. (~\cite{lagrange08}), it is most likely a carbon-rich planet also because it is located in a carbon-rich environment. Carbon planets can also form in a local area enriched in C or depleted in H$_{2}$O in an otherwise solar abundance protoplanetary disk (Seager et al. \cite{seager07}). Lodders (\cite{lodders04}) suggested that the planetary
embryo that grew into Jupiter may have formed in such a locally carbon-rich area, and that Jupiter's embryo was a carbon planet. In this paper, we also consider the carbon model for CoRoT-7b, Kepler-9d, and Kepler-10b, in which SiC is the major mantle constituent. The detailed formation scenarios for Earth-mass carbon planets can be learned from Kuchner \& Seager (\cite{kuchner06}), the latest research about the carbon-rich giant planets can be found in Madhusudhan et al. (\cite{madhusudhan11}).

\section{Numerical Results}

\subsection{Validation in the Solar System}  

\subsubsection{Planets in the Mass-radius Relationship Diagram}
A mass-radius diagram is a useful tool when inferring a planet's bulk composition. Building on the work done by Zapolsky \& Salpeter (\cite{zapolsky69}), we first consider planets of uniform composition. The positions of CoRoT-7b, Kepler-9d, and Kepler-10b in mass-radius diagram can quickly tell us whether we can model them as solid planets. In Figure 1,
the lines are mass-radius relationship curves for homogeneous planets. From top to bottom the homogeneous planets consist of ice VII, SiC, MgSiO$_{3}$ (enstatite, ab. en), (Mg$_{0.88}$, Fe$_{0.12}$)SiO$_{3}$ (perovskite, ab. pv), Fe($\alpha$), Fe($\varepsilon$), respectively. Here, MgSiO$_{3}$ and (Mg$_{0.88}$, Fe$_{0.12}$)SiO$_{3}$ are usually looked as representative material in mantle. The phase diagram of iron in Valencia et al. (2010) can help us to understand  Fe($\alpha$) and Fe($\varepsilon$) phase of iron (under different pressure scope). CoRoT-7b, Kepler-9d, Kepler-10b and the Earth are marked on the mass-radius diagram. Two notable exoplanets GJ 1214b (Charbonneau et al.~\cite{charbonneau09}) and Kepler-11b (Lissauer et al.~\cite{lissauer11}) are introduced here as a comparison. Their measured masses and radii tell us that Kepler-9d, Kepler-10b and CoRoT-7b are candidates of solid super-Earths. GJ 1214b is above the water ice VII line, this indicates it must contain some light elements such as H and He, most likely it has a substantial gas envelope. Kepler-11b is above the solid SiC line, it should have a substantial water vapour or H/He gas envelope because it is so close to host star (10.3 days, Lissauer et al.~\cite{lissauer11}). As the validation, we also mark some planets and satellites ($M \le 1\;{M_ \oplus }$) in our solar system in Figure 2. The Earth and the Venus fall between silicate and iron line, but close to silicate line, it implies they contain more rock composition than iron. Ganymede is found between silicate and the water ice VII line, which indicates it must have significant amounts of ice composition. The Moon just lies on the silicate line, it agrees with the fact that the Moon is composed almost entirely of rock with a small iron core (Hu \& Xu \cite{hu08}). Unlike the Moon, Mercury must have a massive iron core.

\subsubsection{Code Test for the Differentiated Planets}
Mass-radius relationship diagram (or one layer model) can only give us a qualitative estimation, but it can not tell us the details of interior structure. To estimate the precision of our code, we calculate the radii of planets in solar system from their total mass and mass fraction published in relevant literatures. On observation, the planet mass and radius uncertainties are typically $5\%$ to $10\%$ (Selsis et al.~\cite{selsis07}), so a robust numerical code must guarantee the error is less than $5\%$. Table 3 are the results we get. We must emphasize that mass fractions of materials differ in literature, which indicates that there are still some uncertainties in the bulk composition of solar system objects. For example, the iron core mass fraction of the Moon is estimated at less than 10$\%$ of its total mass - 3$\%$ is found by Canup \& Asphaug (\cite{canup01}). We use both pv and en phase of silicate to calculate the radii, the aim is to test the model approach only. Beside for Ganymede and Mars (both with the pv mantle), the relative error of all other planets and the Moon is less than $5\%$. Large relative error to Mars is understandable because Mars is relatively small, so pv is less suitable. Mercury is small too, but it has a large iron core, so pv or en can not cause large discrepancies. The relatively large error ($6.73\%$) of Ganymede (with pv mantle) may be caused by the relatively low pressures in the deep interior of Ganymede (Sohl et al. ~\cite{sohl02}), and therefore pv is not very suitable, either. The calculated Earth radius are within 101km (only $1.59\%$ error) of the actual Earth, this means that pv is the predominant mantle mineralogical phase in Earth.
\begin{table}
\begin{center}
\caption[]{Relative Error Between Calculated Radii and Actual Radii.}\label{Tab:Rerror}

\begin{tabular}{cccc}
  \hline\noalign{\smallskip}
{\bf {Celestial Body Radius (km)}$^{(a)}$} & {\bf {Assumed Composition}$^{(b)}$} & {\bf {Model Radius (km)}} & {\bf {Relative Error ($\%$})}  \\
  \hline\noalign{\smallskip}
 {Mercury} & {60$\%$Fe + 40$\%$en} & {2474.43} & {1.40}   \\   
 {2439.70} & {60$\%$Fe + 40$\%$pv} & {2344.78} & {3.80}  \\   
 \hline
 {Venus} & {27.9$\%$Fe + 72.1$\%$en} & {6283.54} & {3.83}  \\   
 {6051.80} & {27.9$\%$Fe + 72.1$\%$pv} & {5862.68} & {3.13}  \\   
 \hline
 {Moon} & {3$\%$Fe + 97$\%$en} & {1745.05} & {0.43}  \\   
 {1737.50} & {3$\%$Fe + 97$\%$pv} & {1652.37} & {4.90}  \\   
 \hline
 {Earth} & {32.5$\%$Fe + 67.5$\%$en} & {6589.91} & {3.40}  \\   
 {6371.00} & {32.5$\%$Fe + 67.5$\%$pv} & {6269.81} & {1.59}  \\   
 \hline
 {Mars} & {20$\%$Fe + 80$\%$en} & {3409.08} & {0.58}  \\   
 {3389.50} & {20$\%$Fe + 80$\%$pv} & {3149.51} & {7.00}  \\   
\hline
 {Ganymede} & {6.5$\%$Fe + 48.5$\%$en + 45$\%$ice VII} & {2525.19} & {4.02}  \\   
 {2631.20} & {6.5$\%$Fe + 48.5$\%$pv + 45$\%$ice VII} & {2454.14} & {6.73}  \\   
  \noalign{\smallskip}\hline
\end{tabular}
\begin{minipage}{13cm}
{\footnotesize  $^{(a)}$ http://ssd.jpl.nasa.gov/. \\
$^{(b)}$ References: Mercury, Solomon (\cite{solomon03}); Venus, Zharkov (\cite{zharkov83}); Earth, Ganymede, Seager et al. (\cite{seager07});
                     Mars, Rivoldini et al. (\cite{rivoldini11}).}
\end{minipage}
\end{center}
\end{table}

\label{sect:results}

\subsection{The Results for Three Exoplanets}

We examine the interior composition of Kepler-10b, Kepler-9d and CoRoT-7b under the assumption of an iron core covered by a mantle composed of (Mg$_{0.88}$, Fe$_{0.12}$)SiO$_{3}$ (silicate model, similar to Earth's mantle) or SiC (carbon model). When considering two-layer models, the measured mass and radius uniquely determine the
planet's composition. The core mass fraction as a function of planet radius for the three planets are displayed in Figure 3-5. The solid red
line denotes the fraction of planet's mass in its iron core according to an errorless planetary mass, while the
green, white, and red shaded regions in Figure 3-4 delimit the 1$\sigma$, 2$\sigma$, and 3$\sigma$
error bars on ${M_p}$, respectively.
In Figure 5, the solid red line denotes the iron core mass fraction for $M_{p}=5.25\;M_{\oplus}$ (middle mass between 3.5-7.0 $M_{\oplus}$).
The gray regions in Figure 5 delimit the mass range we considered. The measured planet radius and its 1$\sigma$ error bars are denoted by solid blue and the dashed black
vertical lines, respectively.

Interior structure models can strengthen the observational constraints on a planet's mass and radius (Rogers et al. \cite{rogers10}). For example, with the assumption
that planets do not have a significant water or gas layer,
some of the mass-radius pairs within ${M_p} \pm 1{\sigma _M}$ and ${R_p} \pm 1{\sigma _R}$
(including the 0$\sigma$ mass-radius pair) can be ruled out because they correspond to bulk densities lower than a pure
silicate planet. These excluded mass-radius pairs would necessitate water (or some other component lighter than silicate). In Figure 3-5, all mass-radius pairs in the 1$\sigma$ error bar (both mass and radius) compose an area. Here, we call them `effective area'.

To CoRoT-7b, if it has an Earth-like silicate mantle, its radius must be less than the measured value (1.68 $R_{\oplus}$) and its mass has to be larger than the measured mass (4.8 $M_{\oplus}$) within 1$\sigma$ error bars, which implies an iron core mass fraction of less than $\sim 27\%$. Within its observational value, its radius is unlikely less than the reliable value if it formed in the carbon rich environment and has a Moon-like composition (where the iron content is below $10\%$ in carbon model). Its iron core is not heavier than $\sim 70\%$ of its total mass even at 3$\sigma$ error bar (see carbon model in Figure 3), it is the largest value for both models. The silicate model of CoRoT-7b with an Earth-like iron core mass fraction ($32.5\%$) is not consistent with the measured mass and radius within 1$\sigma$. If its mass is ascertained with in 2$\sigma$ or 3$\sigma$ error bar (in the range got by Hatzes et al. \cite{hatzes11}), CoRoT-7b can be consistent with an Earth-like composition. From Figure 3, we can clearly see that the `effective area' in the carbon model is much larger than for the silicate model.

Kepler-10b tells us other information. For its carbon model, there is a lower iron mass fraction limit (about $41\%$) within 1$\sigma$ error bar (see carbon model in Figure 4). Kepler10 (7.4-16.4 Ga) is older than CoRoT-7 (1.2-2.3 Ga), if CoRoT-7b is differentiated as we assumed, Kepler-10b is more likely to be differentiated. This means that Kepler-10b may have a larger iron core. Our simulation indicates Kepler-10b is denser than CoRoT-7b, but the detailed calculation can tell us more: if CoRoT-7b has a silicate mantle, whereas Kepler-10b formed under carbon-rich circumstances, the upper core mass fraction limit of CoRoT-7b ($\sim 27\%$) is below the lower core mass fraction limit ($41\%$ mentioned above) of Kepler-10b! To Kepler-10b's carbon model, the upper limit of the iron core mass fraction is $\sim 84\%$ (see carbon model in Figure 4) within 1$\sigma$ error bar, this value is the largest in its two models and is larger as the suggested value of Mercury ($\sim 60\%$) -- the densest planet in solar system. In Figure 6, we give the histograms of iron content for CoRoT-7b and Kepler-10b. The lower value we take comes from the planet formation scenario which we will discuss later. To Kepler-10b, silicate model is compatible with an Earth-like composition within the 1$\sigma$ error bar, the mass is about 3.27 $M_\oplus$ (lower limit of the observational value). The carbon model of Kepler-10b is not compatible with an Earth-like core/mantle mass ratio (32.5$\%$ in core).

Kepler-9d is shown in Figure 5, the gray band denotes the planetary mass range we are considering. If the maximum mantle-stripping theory is true, silicate model in Figure 5 tells us that Kepler-9d's radius can not be larger than $\sim $1.75$R_{\oplus}$. In detail, if it has an Earth-like core/mantle mass ratio, its radius is no more than $\sim $1.7 $R_{\oplus}$ with silicate model. Despite the mass of a transiting-only super-Earth is unknown, model simulations give constraints on their possible masses. For example, if Kepler-9d has an Earth-like mantle composition (silicate model in Figure 5), it has a lower mass limit according to its smallest possible radius, which is larger than the lower mass limit (3.5 $M_ \oplus$) discussed in Holman et al. (\cite{holman10}). An Earth-like core mass fraction and the lower limit of observational radius ($\sim $1.50 $R_{\oplus}$) yield a total mass of $\sim$4.1 $M_ \oplus$. In the mass range discussed in this paper, the carbon model of Kepler-9d is compatible with an Earth-like core/mantle mass ratio in a large scope of radius.

Other information can be read from the Figure 3-5. For a given mass, the carbon model yields a larger radius than the silicate model. If the core is not pure iron but also contains a light element such as sulfur (in the Earth's core), all of the slantwise lines will sway to the right a little more (having no effect on the radius at a core mass fraction of 0.0). This indicates that the core mass fraction at a specified planetary radius will be larger. Finally, we calculated the possible interior structure of CoRoT-7b and Kepler-10b using their presumable masses (Figure 7-8). To CoRoT-7b, we take an Earth-like core/mantle mass ratio (core mass is 32.5$\%$). From Figure 7, silicate model of CoRoT-7b can be ruled out because the calculated radius is outside the observational 1$\sigma$ error bar. To Kepler-10b, we take a Mercury-like core/mantle mass ratio (core mass is 60 $\%$). In Figure 8, Kepler-10b is compatible with a Mercury-like core/mantle mass ratio with its reliable mass for silicate model.

\section{Discussion and Conclusion}
\label{sect:summary}
Two recently discovered super-Earths, Kepler-9d and Kepler-10b, and previously found CoRoT-7b are discussed in this paper. Because of their intense irradiation and small size, they are possibly absent of gas envelopes. Under the assumption of that they are mainly composed by refractory materials, their internal structures are deduced using suitable equation of state of the materials. The silicate and the carbon model are discussed in details. By analyzing the numerical results, we find some of the mass-radius pairs within ${M_p} \pm 1{\sigma _M}$ and ${R_p} \pm 1{\sigma _R}$ can be ruled out. Therefore, our interior models sharpen the observational constraints on their masses and radii. For CoRoT-7b, an Earth-like composition with a silicate (pv) mantle is less likely within both the mass and the radius 1$\sigma$ error bars, but a Mars-like composition may be suitable. To Kepler-10b's two models, a Mercury-like composition having $60\%$ of its mass in iron core is consistent with the measured mass and radius within 1$\sigma$ error bar. There is an upper and lower limit for the iron core mass fraction of every planet. To CoRoT-7b, the upper iron core mass fraction limits are $27\%$ (silicate model) and $56\%$ (carbon model). For Kepler-10b, the silicate model have an upper limit of about $75\%$ and its iron content is approximately $41\% \sim 84\%$ for the carbon model. Transiting-only exoplanet like Kepler-9d can also be studied, for example, its radius is no more than 1.7 $R_{\oplus}$ in silicate model. The lower mass limit (3.5 $M_ \oplus$ we adopted) is also ruled out in its silicate model.

Our quantitative calculation tells us Kepler-10b may be a Mercury-like planet. CoRoT-7b and Kepler-10b share the similar features: their radii, their period, the type of host star and perhaps their mass. Why is Kepler-10b denser than CoRoT-7b? Their different ages may give us some answers. To
the anomalously dense Mercury in solar system, there are three competing viewpoints (Solomon \cite{solomon03}):
1) the solar nebula caused aerodynamic drag on the particles from which Mercury was accreting, the iron and silicate have different response to this drag, so the lighter particles were lost from the accreting material at the onset of accretion. 2) it surface rock could have been vaporized by strong radiation from a hot nebula and have been carried away by the solar wind, 3) much of the original crust and mantle have been stripped away by a giant impact caused by a planetesimal. Because of lacking detailed observational data, which hypothesis is true can not be determined. However, each hypothesis predicts a different surface composition. So two promising space missions, underway MESSENGER (Solomon et al. \cite{solomon01}, NASA) and the upcoming  BepiColombo (Grard et al. \cite{grard01}, ESA), could bring us the answer. The second and third explanations invoke processes late in the planetary accretion process, after the protoplanet had differentiated silicate mantle from metal core. If CoRoT-7b and Kepler-10b were born in a similar environment, Kepler-10b's abnormal iron content can be naturally associated with its older age. We think the postnatal alteration is more likely.

The lower limit of iron content of a super-Earth is an interesting question because a pure silicate super-Earth (iron mass fraction is zero) is unimaginable under differentiation hypothesis. Under physically plausible conditions analogous to solar system, planet formation will lead to differentiated super-Earth with an iron core covered by a silicate mantle, with the proportions of each determined by the local Si/Fe ratio (Grasset et al. \cite{grasset09}). The only way to significantly increase the mean density of a planet requires removal of an extended part of its silicate mantle. Marcus et al. (2010) proposed an efficient method -- giant impacts between super-Earths. On the other hand, they prefer a lower limit of iron content of $33\%$ by considering the standard cosmic abundances (see also Valencia et al. \cite{valencia07}). The initial Fe/Si ratio (average value) is used to give a lower limit for the iron content of super-Earth. But, in practice, formation progress must be influenced by other factors (such as the survival competition between planets) because some planets in solar system have an iron content lower than this `lower limit'. We think that referring to the iron content in chondrites is a sound choice. L chondrites have lower total iron content, which is about $10\%$ (Chen et al. \cite{chen94}). L chondrites are the largest group in the ordinary chondrites (40$\%$, http://www.nhm.ac.uk/). Therefore, we use $10\%$ as the lower limit for silicate model. Regarding Figure 6, if there is a calculated lower limit of the iron content, we will use it, otherwise we use $10\%$ (only for silicate model).

Our main conclusions are summarized as follows: 1) The iron core of CoRoT-7b is not more than $27\%$ of its total mass within 1 $\sigma$ mass-radius error bars, so an Earth-like composition is less likely, but its carbon rich model can be compatible with an Earth-like core/mantle mass fraction; 2) Kepler-10b is more likely with a Mercury-like composition, its old age implies that its high iron content may be a result of strong solar wind or giant impact; 3) the transiting-only super-Earth Kepler-9d is also discussed. Combining its possible composition with the formation theory, we can place some constraints on the mass and composition of Kepler-9d. The radius derived from transit method and the mass got by radial velocity detections together with the model simulation can constrain the possible compositions of exoplanets.

Credible conclusions desiderate comprehensive observations. Planetary formation theories, thermal evolution models, and studies of the cosmic abundance of elements can be used to place additional constraints on a planet's interior composition. Besides the temperature-independent model used in this paper, more complicated models of solid exoplanets can be found in other literatures. For example, Valencia et al. (\cite{valencia06}) and Sotin et al. (\cite{sotin07}) considered the temperature profile and mixture of materials in the mantle or core. Wagner et al. (\cite{wagner11}) used material laws in the infinite pressure limit to improve the model of solid exoplanets. In the near future, high-sensitivity spectroscopic transit observations of these
exoplanets should constrain the compositions of the
evaporating flow and therefore allow us to distinguish between rocky or gas-rich planets.
In particular for exoplanets, without any opportunity to in situ composition measurements
and gravitational moment measurements from spacecraft flybys, we will be permanently limited in what we can infer about the interior composition from their observational mass and radius (Rogers \& Seager~\cite{rogers10}).

\begin{acknowledgements}
Firstly, we want to thank the anonymous referees for constructive comments and suggestions which improved the paper a lot. This work was supported by the National Natural Science Foundation of China (Nos.10833001 and 10925313), Ph.D training grant of China (20090091110002), and Fundamental Research Funds for the Central Universities (1112020102). Gong Yan-Xiang
also acknowledge support form the Shandong Provincial Natural Science Foundation, China (ZR2010AQ023). We also thank Dr. Zeng Li from Harvard University for his communication with us about the interior structure of close-in exoplanets.
\end{acknowledgements}

\begin{figure}
\centering
\includegraphics[width=15cm, height=12cm]{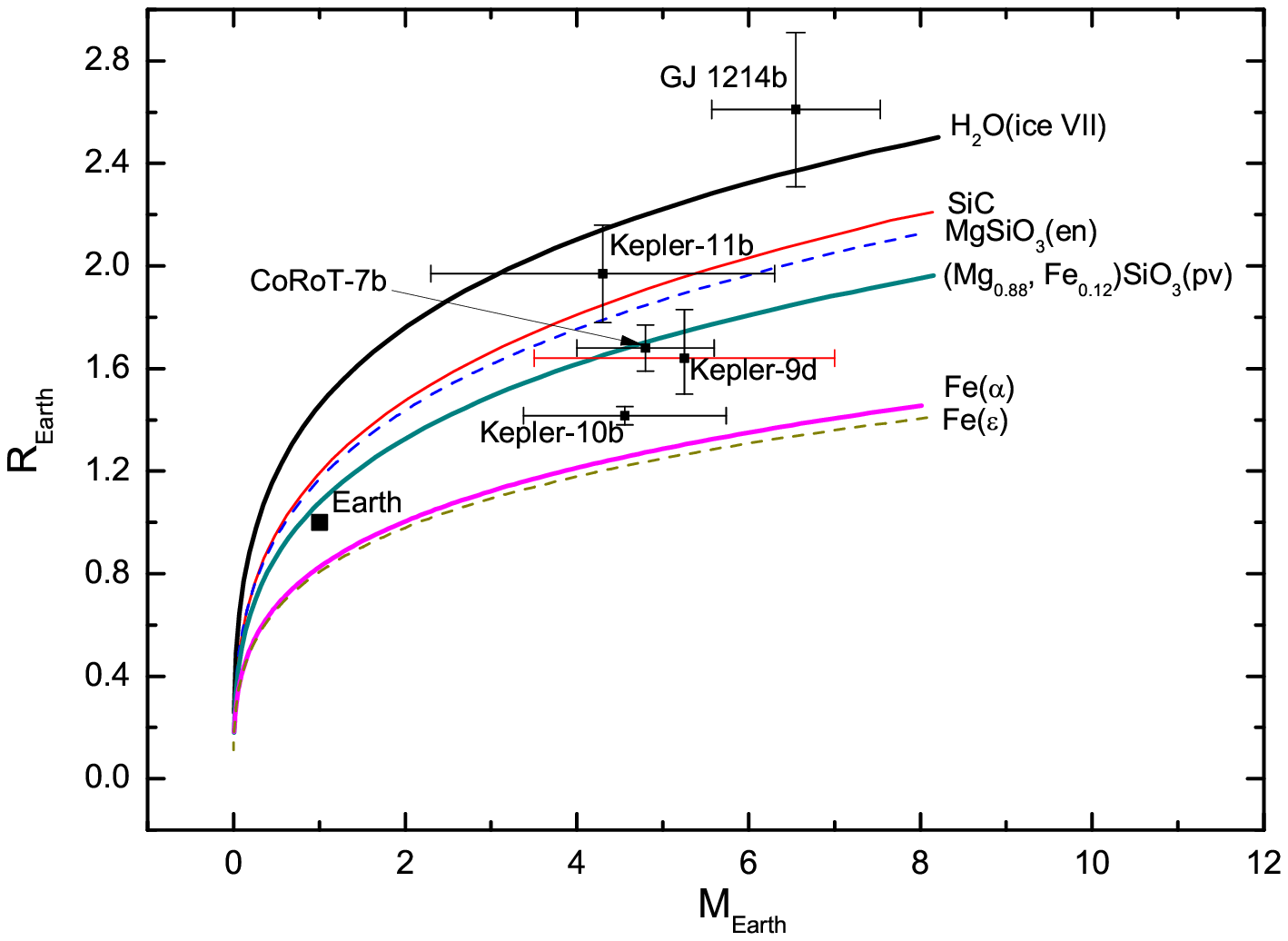}
\caption{Mass-radius relationship diagram for ${M_p} \le 10\;{M_ \oplus}$. The solid or dashed lines are homogeneous planets. Five exoplanets and the Earth are marked on the diagram. The mass range of Kepler-9d (red error bar) is taken from Holman et al. (\cite{holman10}). The locations of CoRoT-7b, Kepler-9d and Kepler-10b indicate they can be described as solid planets.}
\label{Fig:mrr1}
\end{figure}

\begin{figure}
\centering
\includegraphics[width=14cm, height=12cm]{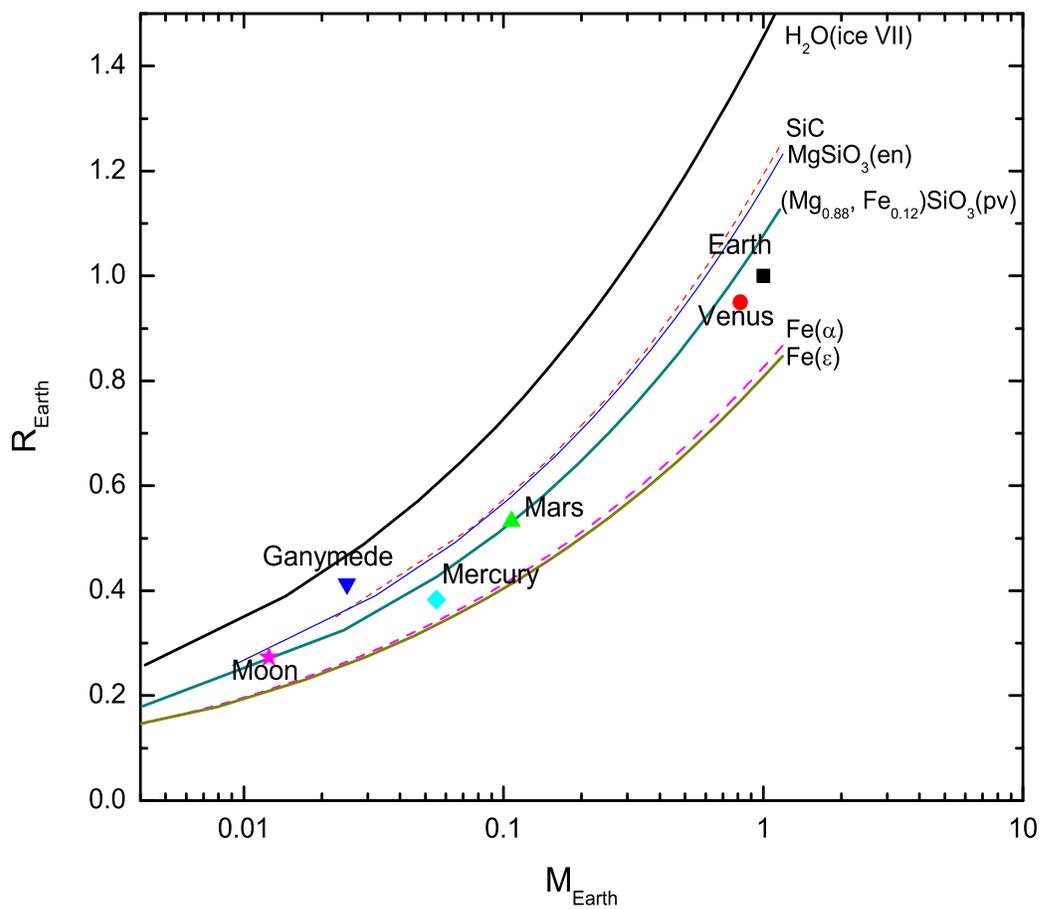}
\caption{Mass-radius relationship diagram for ${M_p} \le 1\;{M_ \oplus }$. The solid or dashed lines are homogeneous planets. Some planets and satellites in solar system are shown in it. Their major constituents can be estimated using this mass-radius diagram.}
\label{Fig:mrr2}
\end{figure}

\begin{figure}
\centering
\includegraphics[width=10cm, height=12cm]{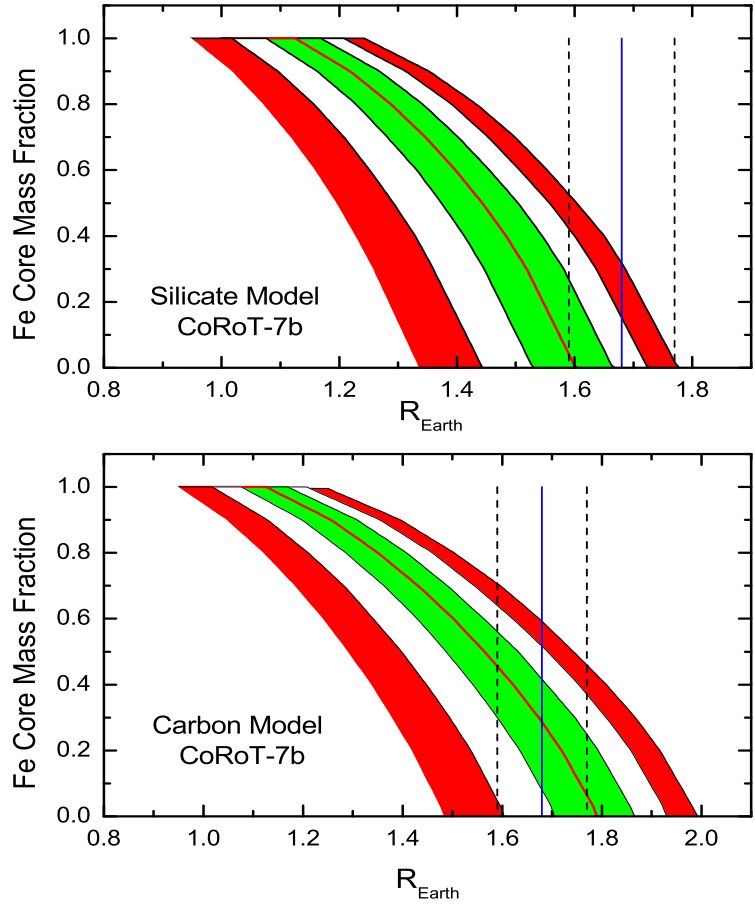}
\caption{Core mass fraction as a function of planetary radius for CoRoT-7b. Silicate model ((Mg$_{0.88}$, Fe$_{0.12}$)SiO$_{3}$) and carbon model (SiC) are discussed. The planetary mass is $M = 4.8 \pm 0.8\;{M_ \oplus }$. The green, white and red regions denote the core mass fractions obtained by varying its mass within 1$\sigma$, 2$\sigma$ and 3$\sigma$ error bars. The dashed black vertical lines delimit the measured radius $R = 1.68\pm 0.09\;{R_ \oplus }$. The Silicate model of CoRoT-7b had been discussed by Rogers \& Seager (\cite{rogers10}) and Valencia et al. (\cite{valencia10}). We redo it for the comparison with its carbon model.}
\label{Fig:cort7b}
\end{figure}

\begin{figure}
\centering
\includegraphics[width=10cm, height=12cm]{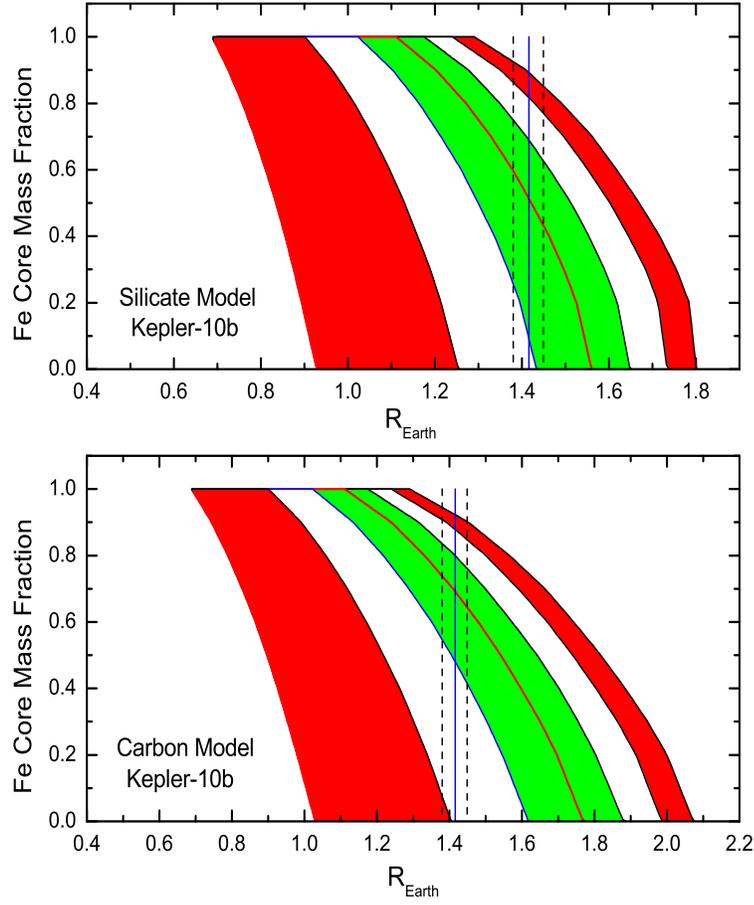}
\caption{Kepler-10b core mass fraction as a function of planetary radius. Silicate model ((Mg$_{0.88}$, Fe$_{0.12}$)SiO$_{3}$) and carbon model (SiC) are discussed. The lines and areas are similar to Fig.3. The planetary mass is $M = 4.56_{- 1.29}^{ + 1.17}\;{M_ \oplus }$ and the measured radius is $R = 1.416_{ - 0.036}^{ + 0.033}\;{R_ \oplus }$. $\Delta M = \pm 1.29\;M_\oplus$ is used for 2$\sigma$ and 3$\sigma$ error bars of its mass.}
\label{Fig:kepler10b}
\end{figure}

\begin{figure}
\centering
\includegraphics[width=10cm, height=12cm]{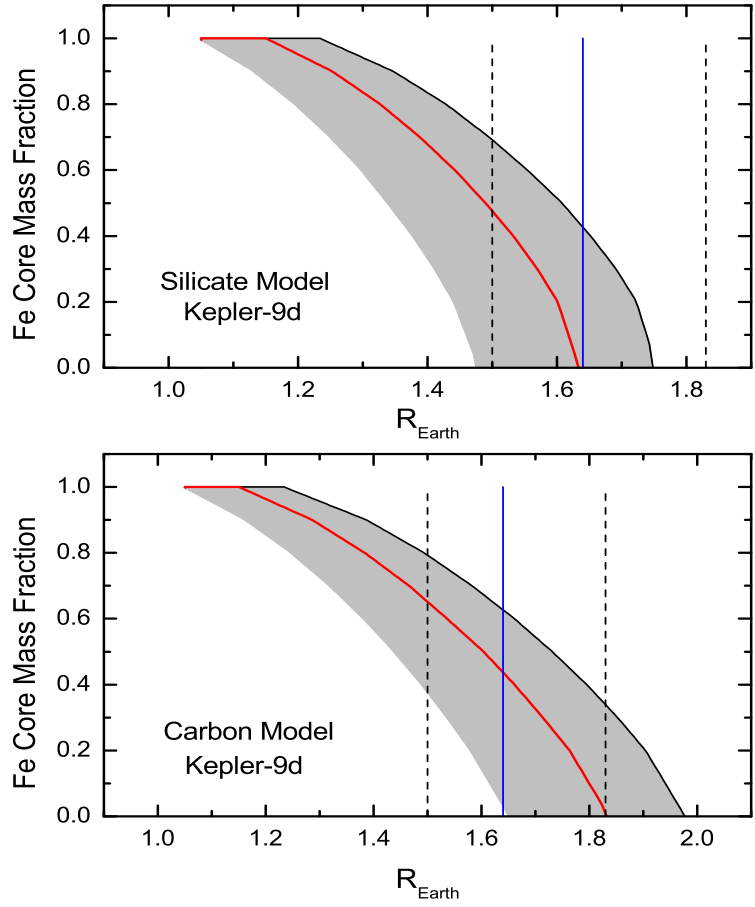}
\caption{Kepler-9d core mass fraction as a function of planetary radius. Silicate model ((Mg$_{0.88}$, Fe$_{0.12}$)SiO$_{3}$) and carbon model (SiC) are discussed. The measured radius is $R = 1.64_{ - 0.14}^{ + 0.19}\;{R_ \oplus }$. The gray region denotes the core mass fractions obtained by varying the mass within the mass range we considered, $M = 5.25 \pm 1.75\;{M_ \oplus }$.}
\label{Fig:kepler9d}
\end{figure}

\begin{figure}
\centering
\includegraphics[width=10cm, height=8cm]{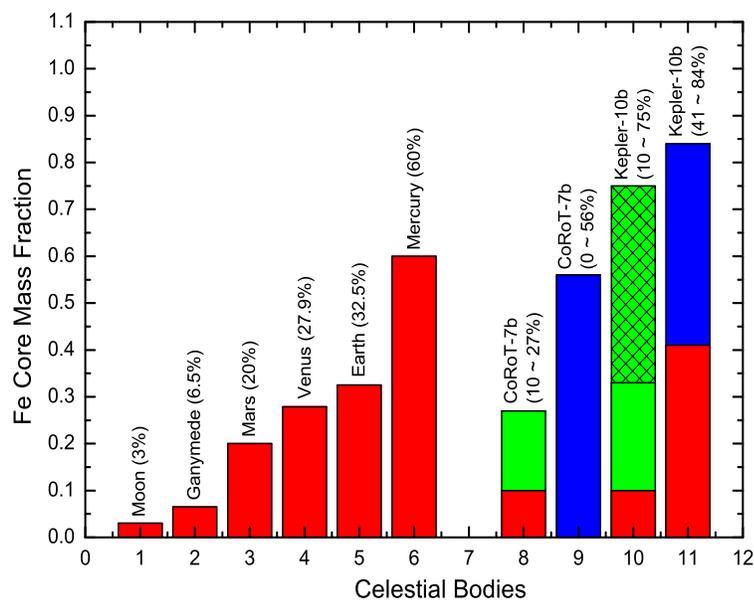}
\caption{The histograms of iron content for CoRoT-7b, Kepler-10b and other celestial bodies mentioned in our work. The corresponding references are the same in Table 3. The green part denotes the upper and lower limit for silicate model, the blue part denotes similar limit for carbon model (see details in Section 4). To the grid range on the histogram of Kepler-10b (silicate model), the {\it lower} limit ($33\%$) adopted by Marcus et al. (\cite{marcus10}) is used. Obviously, from a statistical point of view, CoRoT-7b is compatible with a Mars-like core/mantle mass fraction, whereas Kepler-10b is compatible with a Mercury-like composition. }
\label{Fig:k9dden}
\end{figure}

\begin{figure}
\centering
\includegraphics[width=12cm, height=10cm]{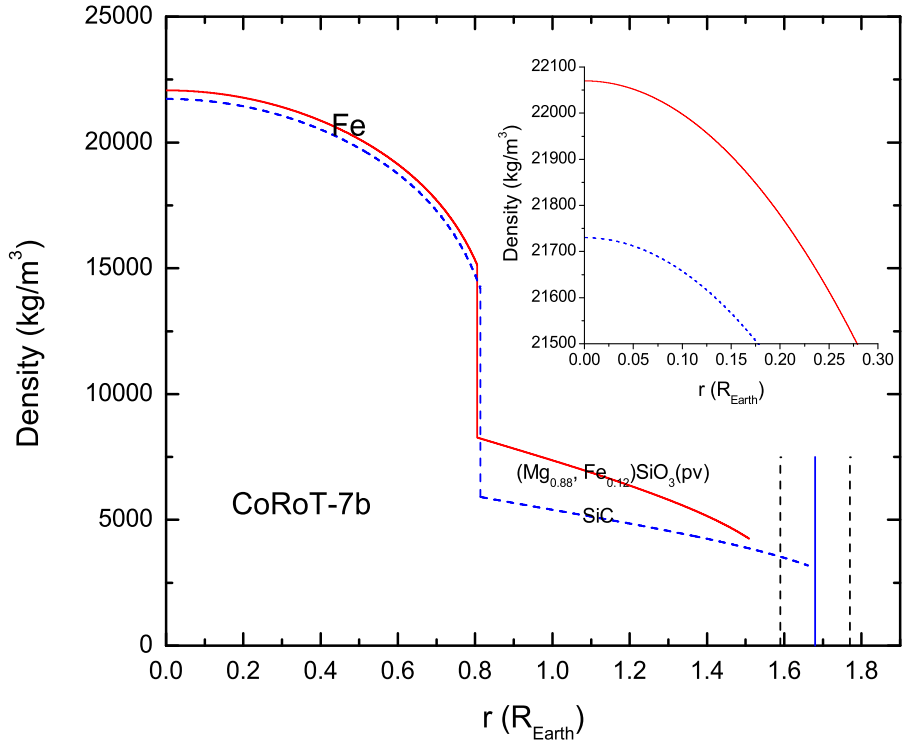}
\caption{The interior structure of CoRoT-7b. We use the reliable mass $M = 4.8\;{M_ \oplus }$ with an Earth-like core/mantle mass ratio (iron core is 32.5$\%$ ). Two mantle compositions are considered, the red line is for (Mg$_{0.88}$, Fe$_{0.12}$)SiO$_{3}$ and the blue line is for SiC. Its radius scope $R = 1.68\pm 0.09\;{R_ \oplus }$ is shown using vertical lines. Silicate model can be ruled out because the corresponding radius is out the 1$\sigma$ error bar for its reliable mass.}
\label{Fig:c7bden}
\end{figure}

\begin{figure}
\centering
\includegraphics[width=12cm, height=10cm]{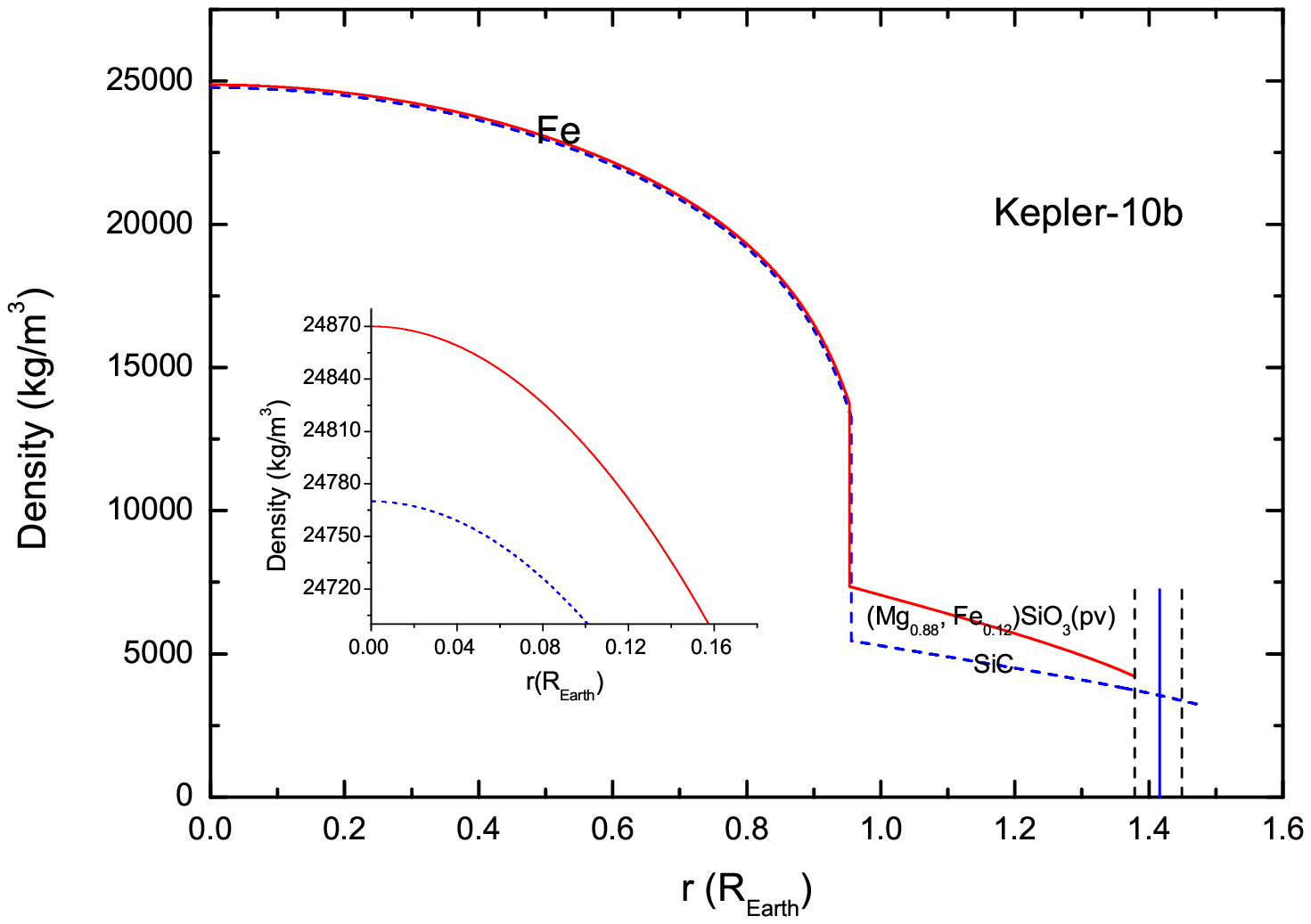}
\caption{The interior structure of Kepler-10b. We use a Mercury-like core/mantle mass ratio (iron core is 60$\%$ ) and its reliable mass $M = 4.56\;{M_ \oplus }$. Two mantle compositions are considered, the red line is for (Mg$_{0.88}$, Fe$_{0.12}$)SiO$_{3}$ and the blue line is for SiC. Its radius scope $R = 1.416_{ - 0.036}^{ + 0.033}\;{R_ \oplus }$ is shown using vertical lines. Carbon model can be ruled out because the corresponding radius is out the 1$\sigma$ error bar for its reliable mass.}
\label{Fig:k10bden}
\end{figure}

\label{lastpage}

\newpage
\begin{thebibliography}{99}

\bibitem[2000]{ahrens00} Ahrens, T. J. 2000, Mineral Physics and Crystallography: A Handbook of Physical Constants (Washington, DC: AGU)
\bibitem[1989]{alexsandrov89} Alexsandrov, I. V., Gocharov, A. F., et al. 1989, J. Exp. Theor. Phys. Lett., 50, 127
\bibitem[2001]{anderson01} Anderson, O. L., Dubrovinsky, L., Saxena, S. K., \& LeBihan, T. 2001, Geophys. Res. Lett., 28, 399
\bibitem[2011]{batalha11} Batalha, N. M., Borucki, W. J., Bryson, S. T., et al. 2011, ApJ, 729, 27
\bibitem[2011]{boisse11} Boisse, I., Bouchy, F., H\'ebrard, G., et al. 2011, A\&A, 528, A4
\bibitem[2003]{bord03} Bord\'e, P., Rouan, D., \& L\'eger, A. 2003, A\&A, 405, 1137
\bibitem[2011]{borucki11} Borucki, W. J., Koch, D. J., Basri, G., et al. 2011, ApJ, 736, 19
\bibitem[2010]{bruntt10} Bruntt, H., Deleuil, M., Fridlund, M., et al. 2010, A\&A, 519, A51
\bibitem[2001]{canup01} Canup, R. M., \& Asphaug, E. 2001, Nature, 412, 708
\bibitem[2009]{charbonneau09} Charbonneau, D., Berta, Z. K., Irwin, J., et al. 2009, Nature, 462, 08679
\bibitem[1994]{chen94} Chen, D. G., Zhi, X. C., \& Yang, H. T. 1994,  Geochemistry (Hefei: China Science and Technology University Press)
\bibitem[2011]{ferraz11} Ferraz-Mello, S., Tadeu dos Santos, M., Beaug\'e, C., et al. 2011, A\&A, 531, A161
\bibitem[2007]{fortney07} Fortney, J. J., Marley, M. S., \& Barnes, J. W. 2007, ApJ, 659, 1661
\bibitem[2000]{gaidos00} Gaidos, E. J. 2000, Icarus, 146, 637
\bibitem[2001]{grard01} Grard, R., \& Balogh, A. 2001, Planet. Space Sci., 49, 1395
\bibitem[2009]{grasset09} Grasset, O., Schneider, J., Sotin, C. 2009, ApJ, 693, 722
\bibitem[2011]{hatzes11} Hatzes, A. P., Fridlund, M. 2011, arXiv:1105.3372
\bibitem[1987]{hemley87} Hemley, R. J., Jephcoat, A. P., Mao, H. K., et al. 1987, Nature, 330, 737
\bibitem[2010]{holman10} Holman, M. J., Fabrycky, D. C., Ragozzine, D., et al. 2010, Science, 330, 51
\bibitem[2008]{hu08} Hu, Z. W., \& Xu, W. B. 2008, Planetary Science (Beijing: Science Press), 341
\bibitem[2010]{jackson10} Jackson, B., Miller, N., Barnes, R., Raymond, S. N., Fortney, J. J., \& Greenberg, R. 2010, MNRAS, 407, 910
\bibitem[1987]{knittle87} Knittle, E., \& Jeanloz, R. 1987, Science, 235, 668
\bibitem[2006]{kuchner06} Kuchner, M., \& Seager, S. 2006, arXiv:astro-ph/0504214
\bibitem[2008]{lagrange08} Lagrange, A. -M., Gratadour, D., Chauvin, G., et al. 2008, A\&A, 493, L21
\bibitem[1975]{larimer75}  Larimer, J. W. 1975, Geochim. Cosmochim. Acta, 39, 389
\bibitem[2011]{leger11} L\'eger, A., Grasset, O., Fegley, B., et al. 2011, Icarus, 213, 1
\bibitem[2009]{leger09} L\'eger, A., Rouan, D., Schneider, J., et al. 2009, A\&A, 506, 287
\bibitem[2004]{leger04} L\'eger, A., Selsis, F., Sotin, C., et al. 2004, Icarus, 169, 499
\bibitem[1974]{lewis74} Lewis, J. S. 1974, Science, 186, 440
\bibitem[2011]{leitzinger11} Leitzinger M., Odert, P., Kulikov, Y. N., et al. 2011, Planet. Space Sci., 10, 1016
\bibitem[2011]{lissauer11} Lissauer, J. J., Fabrycky, D. C., Ford, E. B., et al. 2011, Nature, 470, 53
\bibitem[2004]{lodders04} Lodders, K. 2004, ApJ, 611, 587
\bibitem[2010]{marcus10} Marcus, R. A., Sasselov, D., Hernquist, L., \& Stewart, S. T. 2010, ApJ, 712, L73
\bibitem[2011]{madhusudhan11} Madhusudhan, N., Mousis, O., Johnson, T. V., et al. 2011, ApJ, 743, 191
\bibitem[1977]{olinger77} Olinger, B. 1977, in High Pressure Physics Research: Applications in Geophysics, ed. M. Manghani \& S. Akimoto (New York: Academic), 325
\bibitem[2000]{poirier00} Poirier, J. P. 2000, Introduction to the physics of the Earth's interior (Cambridge: Cambridge University Press), 63
\bibitem[2009]{queloz09} Queloz, D., Bouchy, F., Moutou, C., et al. 2009, A\&A, 506, 303
\bibitem[2011]{rivoldini11} Rivoldini, A., Van Hoolst, T., Verhoeven, O., et al. 2011, Icarus, 213, 451
\bibitem[2006]{roberge06} Roberge, A., Feldman, P. D., Weinberger, A. J., Deleuil, M., \& Bouret, J. C. 2006, Nature, 441, 724
\bibitem[2010]{rogers10} Rogers, L. A., \& Seager, S. 2010, ApJ, 712, 974
\bibitem[1967]{salpeter67} Salpeter, E. E., \& Zapolsky, H. S. 1967, Phys. Rev., 158, 876
\bibitem[2007]{seager07} Seager, S., Kuchner, M., Hier-Majumder, C. A., \& Militzer, B. 2007, ApJ, 669, 1279
\bibitem[2007]{selsis07} Selsis, F., Chazelas, B., Bord\'e, P., et al. 2007, Icarus, 191, 453
\bibitem[2002]{sohl02} Sohl, F., Spohn, T., Breuer, D., \& Nagel, K. 2002, Icarus, 157, 104
\bibitem[2001]{solomon01} Solomon, S. C., McNutt, R. L., Gold, R. E., et al. 2001, Planet. Space Sci., 49, 1445
\bibitem[2003]{solomon03} Solomon, S. C. 2003, Earth Planet Sci. Lett., 216, 441
\bibitem[2007]{sotin07} Sotin, C., Grasset, O., \& Mocquet, A. 2007, Icarus, 191, 337
\bibitem[1989]{takahashi89} Takahashi, Y. X., \& Spain, I. L. 1989, Phys. Rev. B, 40, 993
\bibitem[2011]{torres11} Torres, G., Fressin, F., Batalha, N. M., et al. 2011, ApJ, 727, 24
\bibitem[2010]{valencia10} Valencia, D., Guillot, T., \& Nettelmann, N. 2010, A\&A, 516, A20
\bibitem[2006]{valencia06} Valencia, D., O'Connell, R. J., \& Sasselov, D. D. 2006, Icarus, 181, 545
\bibitem[2007]{valencia07} Valencia, D., Sasselov, D. D., \& O'Connell, R. J. 2007, ApJ, 665, 1413
\bibitem[2011]{wagner11} Wagner, F. W., Sohl, F., Hussmann, H., Grott, M., \& Rauer, H. 2011, Icarus, 214, 366
\bibitem[1992]{wolszczan92} Wolszczan, A., \& Frail, D. A. 1992, Nature, 355, 145
\bibitem[1993]{wood93} Wood, J. A., \& Hashimoto, A. 1993, Geochim. Cosmochim. Acta, 57, 2377
\bibitem[1969]{zapolsky69} Zapolsky, H. S., \& Salpeter, E. E. 1969, ApJ, 158, 809
\bibitem[1983]{zharkov83} Zharkov, V. N. 1983, M\&P, 29, 139
\end{thebibliography}
\end{document}